\documentclass[prd, twocolumn, nofootinbib]{revtex4}

\usepackage{epsfig} 

\newcommand{\beq}{\begin{equation}}
\newcommand{\eeq}{\end{equation}}
\newcommand{\beqa}{\begin{eqnarray}}
\newcommand{\eeqa}{\end{eqnarray}}
\newcommand{\om}{\Omega_m}

\newcommand{\winf}{w_\infty}

\def\fun#1#2{\lower3.6pt\vbox{\baselineskip0pt\lineskip.9pt
  \ialign{$\mathsurround=0pt#1\hfil##\hfil$\crcr#2\crcr\sim\crcr}}}

\begin{document} 

\title{Biased Cosmology: Pivots, Parameters, and Figures of Merit} 
\author{Eric V.\ Linder} 
\affiliation{Berkeley Lab, University of California, Berkeley, CA 94720} 

\begin{abstract} 
In the quest for precision cosmology, one must ensure that the cosmology 
is accurate as well.  We discuss figures of merit for determining from 
observations whether the dark energy is a cosmological constant or 
dynamical, with special attention to the best determined equation of state 
value, at the ``pivot'' or decorrelation redshift.  We show this is not 
necessarily the best lever on testing consistency with the cosmological 
constant, and moreover is subject to bias.  The standard parametrization 
of $w(a)=w_0+w_a(1-a)$ by contrast is quite robust, as tested by 
extensions to higher order parametrizations and modified gravity.  
Combination of complementary probes gives strong immunization against 
inaccurate, but precise, cosmology. 

\end{abstract} 


\date{\today}

\maketitle

\section{Introduction} \label{sec:intro}

Discovery of the acceleration of the cosmic expansion has precipitated 
widespread activity and plans to uncover its nature.  A significant 
first question to answer is whether the responsible physics is consistent 
with a cosmological constant in Einstein's field equations.  Beyond 
this, we seek to know the dynamics of the dark energy, for example 
through the value and variation of the equation of state, or pressure 
to energy density ratio. 

In evaluating the leverage of experiments one needs to put forth 
either a set of benchmark models to distinguish or a model independent 
parametrization that places a measure or distance in model space. 
Given the lack of well motivated benchmarks other than the cosmological 
constant, the second approach is favored in the literature.  The ability 
of an experiment can then be quantified by a figure of merit, ideally 
rooted in distinctions in the fundamental physics.  Recently, the 
DOE-NASA-NSF Dark Energy Task Force (DETF: \cite{detf}) proposed one 
possible figure of merit, involving the area of the equation of state 
uncertainty contour. 

We examine to what extent this suggestion gets to the heart of the 
physics, in \S\ref{sec:merit}.  Moreover, apart from the theoretical 
precision of a parameter constraint, we must be concerned with the 
accuracy of the parametrization -- the possibility of observational 
and interpretational bias of the parameters and the cosmology.  We 
consider this from 
several angles in \S\ref{sec:bias}, treating issues of decorrelated 
parameters, astrophysical evolution and dark energy time variation, 
and beyond Einstein gravity. 
In \S\ref{sec:contour} we return to the question of figure of merit 
and discuss the role of joint confidence contours.

\section{Figure of merit} \label{sec:merit}

How to robustly quantify knowledge of the nature of dark energy is not 
clear in general.  For example, a cosmological constant has a particular 
value of the equation of state $w=-1$ and lacks any variation, $w'\equiv 
dw/d\ln a=0$.  But 
in the phase plane $w'$--$w$ is the optimum measure the area of a 
likelihood confidence level contour, the volume of the overall N-dimensional 
(equation of state plus matter density etc.\ parameters) confidence surface, 
the largest eigenvalue (shortest dimension of the ellipse defined near the 
likelihood maximum), etc.\ (see \cite{hutturaip} for an early discussion 
of some of these)? 

This difficulty persists even if we restrict to the narrow 
question of whether the dark energy is a cosmological constant $\Lambda$ 
or not.  If we strive to optimize the determination of the value $w$ at 
some redshift, obtaining a so-called sweet spot or pivot 
value $w_p$, we can be misled.  Even if we determine the value $w_p$ is -1 
the variation $w'$ is still important since models distinct from $\Lambda$ 
exist where $w_p=-1$, $w'\ne0$.  Indeed in \S\ref{sec:dens} we will show 
that biased parametrization can lead to exactly this situation. 

\begin{figure}[!hbt]
\begin{center} 
\psfig{file=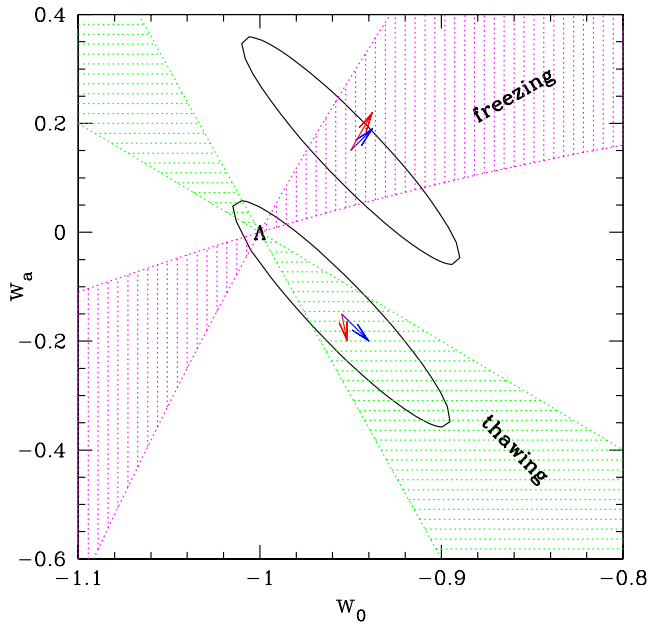,width=3.4in} 
\caption{By using the physical structure of the dark energy phase plane, 
we find optimal quantities for figures of merit for different regions. 
The shaded areas show the freezing and thawing classes \cite{caldlin}; 
they intersect at the cosmological constant $\Lambda$.  Confidence contours 
from next generation cosmological observations are shown for two dark 
energy models, one in each class, pointing up very different criteria for 
distinguishing them from $\Lambda$ (dark, blue arrows) or from the other 
class (light, red arrows). 
}
\label{fig:pinwheel} 
\end{center} 
\end{figure}

In a completely blank phase space, minimizing the area of the equation 
of state contour, as proposed by the DETF in the gaussian approximation, 
is a good strategy.  However, 
we are not completely 
blind to the structure of the physics;  Fig.~\ref{fig:pinwheel} shows 
distinct, physically motivated regions identified by \cite{caldlin} 
due to the overall dynamics of scalar fields and further generalizations 
including modified gravity \cite{paths}.  Thawing models evolve away 
from the cosmological constant state and possess $1+w<w'<3(1+w)$, 
while freezing models move toward the cosmological constant and today 
lie in $0.5w(1+w)\gtrsim w'>3w(1+w)$.  (In the figure we translate to 
$w_a=-2w'|_{z=1}$ to fit the standard parametrization used in the rest 
of the paper.)  This physical structure suggests 
specific strategies for optimizing our knowledge of dark energy.

While there are benefits to uniform shrinking of error contours (which 
translates to minimizing $\sigma(w_p)\times\sigma(w_a)$ in the gaussian 
approximation), the 
phase space structure shows other approaches can be more valuable. 
For a model in the freezing region, we see that more tightly 
constraining the narrow dimension of the contour, corresponding to the 
largest eigenvalue, is near optimal to distinguish the data from the 
cosmological constant -- and to separate it from the other physical 
class of the thawing behavior.  

However, for a model in the thawing 
region, limiting the long dimension of the contour (the {\it smallest\/} 
eigenvalue), is the key discriminant from $\Lambda$.  Indeed, experiments 
that are only sensitive to an averaged equation of state (such as most 
planned ground based experiments, due to limited redshift baseline) 
have a degeneracy direction squarely 
through the thawing region, making it virtually impossible for these 
to distinguish that half of physics classes from the cosmological constant. 
Beyond $\Lambda$, to maximally separate data pointing to the thawing region 
from freezing models one would seek to constrain the contour along nearly 
the vertical direction. 

The uncertainty $\sigma(w_p)$ in the pivot value $w_p$ (see the next 
section for exactly what is a pivot) corresponds not 
to the short dimension of the contour, but the width of the contour at 
fixed $w'$ (or $w_a$), i.e.\ $w'$ held to its fiducial value and not 
treated as a free parameter.  This gives a horizontal cut slanting 
across the contour, not along the short axis (unless the equation of 
state variables are uncorrelated; see \S\ref{sec:contour} for illustration 
of this and further discussion).  Minimizing the horizontal width of 
the contour only gives the optimal science in the special case when 
the data lie horizontally offset from the cosmological constant -- i.e.\ 
in the ``no model's land'' between the freezing and thawing regions, 
where the scalar field potential is fine tuned to balance the kinetic 
energy just so to give a constant value $w$.\footnote{This region can 
be attained by a frustrated network of topological defects, giving the 
constant value $w=-N/3$ for $N$ the value of the defect dimension.  For 
a cosmic string dominated universe, $w=-1/3$, and for domain walls 
$w=-2/3$, both of which are strongly disfavored by current data 
\cite{spergel06}.} 
Furthermore, horizontal squeezing of the confidence contour is never 
optimal for distinguishing the freezing and thawing classes. 

Table \ref{tab:maxsep} summarizes the optimal characteristics to constrain 
-- ``figures of merit'' -- for greatest science return on the dark energy 
equation of state.  Clearly no one figure of merit is optimal in all 
circumstances.  However, we do not know how to weight the likelihood 
that the true model will lie in a particular region of phase space, so 
it is difficult to look for even a ``best average'' figure of merit, that 
gives the greatest good for the greatest numbers, or greatest insight for 
the greatest likelihood of cases.  We discuss this further in 
\S\ref{sec:contour}. 

\begin{table}
\caption{Figures of merit vary for different circumstances.  
Case denotes the region where the true universe lies (blank meaning all 
points in phase space are equivalent).  Goal denotes the science 
objective, e.g.\ distinction from $\Lambda$ or between thawing and 
freezing classes. \label{tab:maxsep}} 
\begin{ruledtabular} 
\begin{tabular}{c|c|c}
\qquad Case\qquad{} &\qquad\quad Goal\qquad\quad{} & \quad Figure of Merit\quad{} \\ 
\hline 
Blank & Anything & Area \\ 
Thawing & $\Lambda$ & Long axis \\ 
Thawing & vs.\ Freezing & $\sim w_a$ \\ 
Freezing & $\Lambda$ & Short axis \\ 
Freezing & vs.\ Thawing & $\sim$ Short axis \\ 
Defect & $\Lambda$ & $w_p$ 
\label{table:maxsep} 
\end{tabular} 
\end{ruledtabular}
\end{table}

What would be beneficial is the power to significantly alter the 
orientation of the contours at a given fiducial point in the phase 
space.  If we could rotate the contour around a thawing model 90 
degrees then these would be more easily distinguishable from a 
cosmological constant (whereas freezing models would be less so). 
This is completely different from constraining $w_p$, i.e.\ 
minimizing $\sigma(w_p)$, however. 
The orientation, or main degeneracy direction, is related to the 
pivot redshift $z_p$ itself, and is a property of the cosmological probe, 
or combination of probes, used.  Our desired 90 degree rotation 
corresponds to a negative $z_p$.  Practical probes all tend to have 
nearly the same degeneracy direction (see, e.g., \cite{coorayhut}) 
and combining probes averages the direction, so it is very difficult 
to rotate the contours even tens of degrees.  (Note that strong 
lensing does provide a way to obtain $z_p<0$ \cite{linsl}, but 
systematics distort the contour so as to prevent significantly 
useful rotation.) 

To understand the heart of dark energy physics we need high precision 
to constrain the dynamics $w'$--$w$; we have seen that a figure of 
merit for that precision is nontrivial.  But we also must make sure 
that the cosmology fitting and its interpretation are accurate, that 
there is negligible bias.  From Fig.~\ref{fig:pinwheel} we see that a 
shift in the best fit parameters could easily move a thawing model, 
say, to a result appearing as the cosmological constant or a freezing 
model.  Moreover, we must test the parametrization of the phase plane 
itself, to ensure that with a few parameters we robustly capture the 
key physics.  The next section studies these two issues, of bias in 
the observations and in the interpretation.

\section{Bias} \label{sec:bias} 

For model independent treatment of the equation of state (EOS) phase space 
$w'$--$w$, the standard parametrization contains two parameters, describing 
the value and variation of the EOS: 
\beq 
w(a)=w_0+w_a(1-a). \label{eq:w0}
\eeq 
This is well known to be accurate in describing a wide variety of 
physical models \cite{linprl,paths}.  Moreover, \cite{linhut05} 
demonstrated that even next generation observations would be generically 
capable of tightly constraining only two parameters.  However, they 
pointed out that extended or higher order parametrizations could be 
useful in assuring the resistance to bias of the fitted cosmological 
parameters. 

Another possibility for dark energy parametrization is the closely related 
form employing decorrelated variables, 
\beq 
w(a)=w_p+w_a(a_p-a). \label{eq:wp}
\eeq 
Here the pivot redshift \cite{astier,huttur,welal} $z_p=a_p^{-1}-1$ 
is chosen such that $w_p\equiv w(z_p)$ and $w_a$ are uncorrelated. 
Known drawbacks of this parametrization include the dependence 
of $z_p$ and hence $w_p$ on the method of probing the cosmology, the 
specific experiment design, the fiducial model, and the priors employed.  
This dependence 
causes a lack of independent physical meaning of $w_p$, i.e.\ $w_p$ 
from one specific experiment for learning about the universe cannot be 
directly and generally compared to the $w_p$ value from another.  The 
plus side is that $w_p$ can be more precisely determined than $w_0$, 
and indeed is the most precisely determined single value of $w(z)$. 

The question we take up here is whether the $w_p$--$w_a$ parametrization 
is also more {\it accurately\/} determined than $w_0$--$w_a$.  We 
examine their robustness against theoretical and observational biases 
that lead to a shift in the fitted parameters. 

Parameter biases $\delta p_i$ induced from offsets $\Delta O_k$ in the 
observable quantities are calculated using the Fisher formalism, 
where maximizing the likelihood leads to 
\beq
\mathbf{\delta p}=A\,\mathbf{\Delta O}=\left(U^T C^{-1}U\right)^{-1} U^T 
C^{-1} \mathbf{\Delta O}, 
\eeq
where $\mathbf{O}$ is the vector of expected observations, $C$ is the 
covariance matrix of observational errors, and $U=\partial O/\partial 
\mathbf{p}$.  Put more simply, when the covariance matrix is diagonal, 
\beq 
\delta p_i=(F_{\rm tot}^{-1})_{ij} \sum_k \frac{\partial O_k}{\partial p_j} 
\frac{1}{\sigma_k^2}\,\Delta O_k, \label{eq:dp}
\eeq 
and $O_k$ is the $k$th observable (e.g.\ supernova magnitude at some 
redshift bin), $\Delta O_k$ is the observational quantity offset (due 
to either theory error or systematic measurement error), and $F_{\rm tot}$ 
is the Fisher matrix from all observables.  We consider the cosmological 
probe of the distance-redshift relation for $z=0-1.7$, as from a future 
survey of Type Ia supernovae by the Supernova/Acceleration Probe (SNAP 
\cite{snap}, which of course includes systematics), together with future 
measurement of the distance to the cosmic microwave background (CMB) 
last scattering surface, as from Planck \cite{planck}.

\subsection{Extended time variation} \label{sec:biasb} 

Disallowing any time variation in the dark energy EOS, i.e.\ holding 
the value $w$ constant for all redshifts or fitting only an averaged 
value, not only obscures the physics but even biases the constant or 
averaged value (e.g.\ \cite{maor}).  The $w_a$ parametrization removes 
that difficulty by allowing for variation, $w_a=-dw/da$.  We should 
ensure, however, that there is not a continued hierarchy of bias, 
i.e.\ that neglecting a second derivative term does not bias the 
measured variation $w_a$. 

To some extent, this robustness is already taken care of by the 
physical foundation of the $w_a$ parametrization -- that it was 
designed specifically to model accurately the exact effect of a 
range of dark energy models (unlike the vast majority of alternate 
parametrizations).  Nevertheless, it is worth explicitly demonstrating 
the robustness. 

To do this, we adopt a parametrization with an explicit second 
derivative $d^2w/da^2\ne0$.  We choose Model~3.1 from \cite{linhut05}, 
\beq 
w(a)=w_0+w_a(1-a^b), \label{eq:ab}
\eeq 
again because of its physical foundation as discussed in \cite{linhut05}. 
Now $d^2w/da^2=-b(b-1)w_a a^{b-2}$.  The phase space behavior is 
given by $w'=b(w-\winf)$, where $\winf$ is the high redshift value of 
$w$.  By considering cases corresponding to both 
freezing and thawing models (evolution toward and away from cosmological 
constant behavior, respectively), we can investigate the stability 
of the first derivative only ($w_a$) parametrization. 

\begin{figure}[!hbt]
\begin{center} 
\psfig{file=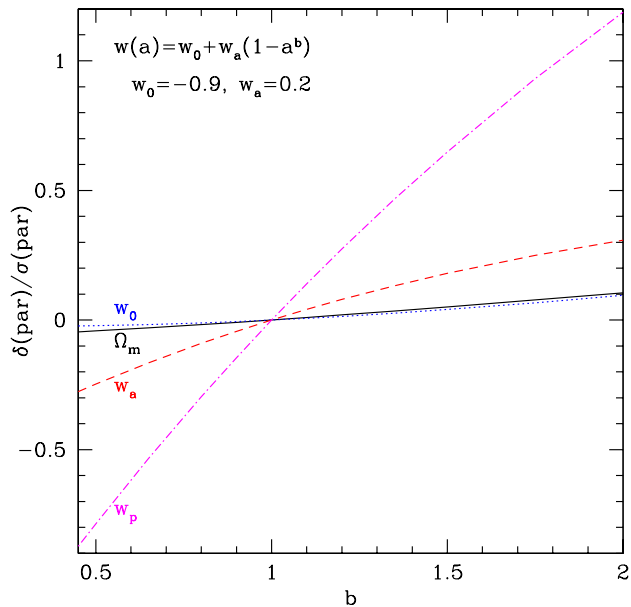,width=3.4in} 
\caption{Dark energy with more complex time variation than the standard 
two parameter model, but interpreted in terms of $w_0$, $w_a$, will bias 
the parameter estimation.  Here we consider observational constraints 
from next generation supernova and CMB measurements on a three parameter 
model in the freezing class.  The $w_0$ and $w_a$ parameters are quite robust, 
nearly unbiased, but the sweet spot, or pivot, value $w_p$ can 
be appreciably misestimated. 
}
\label{fig:biasb2} 
\end{center} 
\end{figure}

First, consider a freezing model with $w_0=-0.9$, $w_a=0.2$.  As we 
change $b$ from its canonical value of unity, but attempt to fit the 
cosmology within the standard parametrization, we will unavoidably 
bias the values of the cosmological parameters.  
Figure~\ref{fig:biasb2} 
shows the fractional bias $\delta p/\sigma(p)$, i.e.\ how many standard 
deviations the parameter $p$ is offset, for joint estimation of 
$\{\om,w_0,w_a\}$ or alternately $\{\om,w_p,w_a\}$ (plus any nuisance 
parameters).  That is, we compare 
the robustness of parametrization (\ref{eq:wp}) vs.\ (\ref{eq:w0}). 
While $w_0$ is essentially unbiased, $w_p$ can suffer bias at a 
significant fraction of the $1\sigma$ statistical uncertainty ($w_a$ and 
$\om$ are negligibly affected).  Note 
that the dark energy model considered here lies within the freezing 
region $w(1+w)\gtrsim w'>3w(1+w)$ for $b=0.45$--1.35. 

Next we consider the thawing 
class of models.  As mentioned in \cite{linhut05}, when the high redshift 
value of the EOS $\winf=-1$, i.e.\ starting from a frozen, cosmological 
constant like state, then the parametrization (\ref{eq:ab}) describes 
a line of slope $b$ in the phase plane: $w'=b(1+w)$, an excellent 
approximation to a wide variety of physics models such as power law 
potentials and pseudo-Nambu Goldstone boson models.  
The thawing region $1+w<w'<3(1+w)$ is obtained for $b=1$--3. 

\begin{figure}[!hbt]
\begin{center} 
\psfig{file=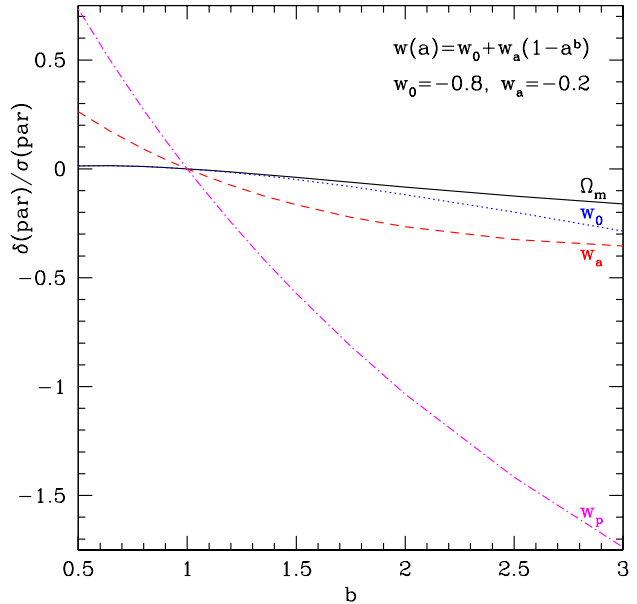,width=3.4in} 
\caption{As Fig.~\ref{fig:biasb2}, but for a model evolving away from 
a high redshift cosmological constant like state, i.e.\ thawing. 
}
\label{fig:biasbm2} 
\end{center} 
\end{figure}

Figure~\ref{fig:biasbm2} illustrates that again the fractional bias 
is well under control for the $w_0$--$w_a$ parametrization, but that 
$w_p$ is subject to large biases.  At the upper end of the thawing 
region, where $b=3$, $w_p$ can be shifted by 1.7$\sigma$, an unacceptable 
bias to the cosmology.  This opens up the prospect of both false positives 
and false negatives, where we either mistakenly think we have confirmed 
the cosmological constant to high precision or incorrectly claim evidence 
for deviation from the cosmological constant despite $\Lambda$ being true. 
In contrast, $w_0$ is misestimated by less than 0.3$\sigma$ (and $w_a$ 
and $\om$ are also cleanly recovered).

So the $w_0$--$w_a$ parametrization is an excellent, fair description 
of the cosmological model even when the dark energy has extended time 
variation, in both the freezing and thawing classes of physics.  These 
classes describe many standard, physically motivated scenarios (see 
\cite{caldlin,paths} for a wide variety of both scalar field and 
modified gravity models).  However one can certainly postulate other 
behaviors with more extreme time variation.  We test the standard 
parametrization to the breaking point in the next section.

\subsection{Rapid transition} \label{sec:biase}

Rapid transitions in dark energy behavior cannot be well fit by the 
simple two parameter model.  In this case, one might require 
four parameters -- asymptotic past and future values, time of transition, 
and rapidity of transition -- as discussed in \cite{linhut05}.  That 
article put forward a solution in Model 4.0, the e-fold model, 
\beq 
w(a)=w_f+\frac{\Delta w}{1+(a/a_t)^{1/\tau}}, 
\eeq
where $w_f$ ($\winf$) is the asymptotic future (past) value, 
$\Delta w=\winf-w_f$, $a_t$ the transition scale factor, and $\tau$ the 
rapidity (note that in \cite{paths}, $\tau$ was the reciprocal of the 
rapidity). 

As the transition becomes more rapid, we expect the two parameter model 
to have greater difficulty fitting this behavior, and hence greater 
bias in the parameter estimation.  Within the e-fold parametrization, 
$\tau$ gives the characteristic e-folding scale of the transition, with 
$\tau=1$ corresponding to a Hubble time at the transition redshift. 
This provides a natural scale, and one might expect that inertia in 
the field evolution could lengthen the transition to $\tau\gtrsim1$.  

\begin{figure}[!hbt]
\begin{center} 
\psfig{file=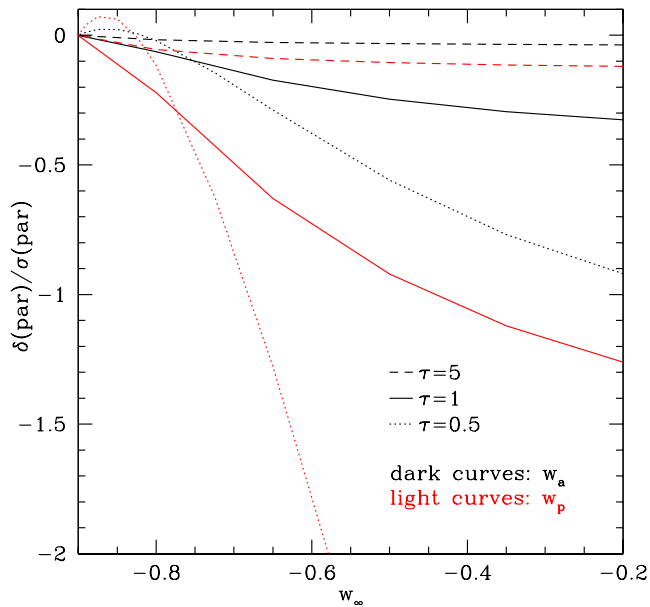,width=3.4in} 
\caption{As Fig.~\ref{fig:biasb2}, but for the e-fold model, as a function 
of $w_\infty=w(z\gg1)$, with $w_0=-0.9$.  When the equation of state 
evolution is rapid (small $\tau$), then estimates of the equation of state 
from the observations are increasingly biased.  The pivot value $w_p$ (light, 
red curves) is most distorted, while $w_a$ (dark, black curves) is fairly 
robust (and $w_0$, not shown, is more so), except for very sharp transitions. 
}
\label{fig:biase} 
\end{center} 
\end{figure}

Figure~\ref{fig:biase} 
shows the bias effects of a cosmology with an e-fold parametrization 
behavior but interpreted in terms of $w_0$--$w_a$ or $w_p$--$w_a$.  As 
expected, the two parameter models are good fits for slow transitions, 
but not for rapid transitions.  The parameter behavior also depends 
on the magnitude of the transition $\Delta w$; here we fix $w_f=-1$ 
and use $\winf=\Delta w-1$ as the independent variable, also taking
$w_0=-0.9$ as fiducial.  Parametrizing in terms of $w_p$ begins to 
be appreciably 
biased for $\tau\lesssim2$, while $w_0$ stays a fair estimator to 
$\tau\gtrsim0.5$.  The bias in $w_a$ tends to be greater than that 
in $w_0$ (and is slightly greater than in $\om$), but is still 
acceptable for $\tau\gtrsim0.7$.

As the limit of an extreme transition we can consider a step function 
in the EOS at some redshift $z_c$.  Even for small steps this can produce 
a large bias in $w_p$, and again the fractional bias in $w_0$ is less. 
Note that \cite{simpsonbridle} found that for an increasing EOS one could 
in fact obtain a bias giving a counterintuitive negative $w_a$.  
While such eigenmode decomposition of the EOS is an elegant approach, 
the Fisher bias formalism equivalently gives the weight or ``redshift 
sensitivity'' functions for parameter fitting (and can handle more 
general situations, as seen in following sections).  
The seemingly paradoxical bias 
result is explained by the degeneracies among the parameter set: while 
$w_a$ may be biased down, $w_0$ or $\om$ would be biased up to compensate. 

The possibility of individual parameters 
flopping around through degeneracies, even to the opposite sign, cautions 
against interpretation relying on a single parameter (as was done with 
constant $w$ in the early days).  
Analogously, it is important not 
to rely on a single probe.  For the exact case considered by 
\cite{simpsonbridle}, the opposite direction paradox goes away if one 
considers both supernovae and CMB distance measurements, as the degeneracy 
is then restricted.

\subsection{Evolving observable} \label{sec:obsevo} 

While we have so far considered the bias due to theory, in terms of 
parametrization, we can apply the same formalism to systematic offsets 
in the measurements.  For supernovae, treatments include the possibility 
of evolving luminosity (e.g.\ \cite{astier,welal,klmm}) and 
gravitational lensing 
amplification bias \cite{holzlinder}.  Effects due to heterogeneous data 
sets were discussed in \cite{linmiq} and systematics in dust extinction 
properties in \cite{lintaup}.  The propagation of systematic errors 
through to cosmological parameter biases is unfortunately less common 
for other cosmological probes (but see \cite{huttak} for a nice treatment 
within weak gravitational lensing, and \cite{mohr} for one instance 
for cluster masses). 

Here we combine the two issues by examining how a measurement systematic 
interacts with the theory parametrization.  We take a hypothetical 
offset in supernova luminosity in terms of the absolute magnitude 
parameter $\Delta M(z)\equiv M(z)-M(0)=\alpha(1-a)$.  This could perhaps arise 
from population drift in supernova environments or progenitor systems 
(without attempting rigorous justification, we note that the offset 
should be bounded as the redshift gets large, especially as the cosmic 
time intervals grow shorter and the diversity of environments decreases).  
We find that the resulting parameter biases can be written as 
\beq 
\frac{\delta p}{\sigma(p)}=X(p)\,\frac{\Delta M(z=1.7)}{0.02}, 
\eeq 
with $X(\om,w_0,w_a,w_p)=-0.55$, -0.14, -0.08, -0.50.  So as before, 
$w_p$ is substantially more biased in units of standard deviation 
than $w_0$.  While $w_a$ is very fairly estimated, now $\om$ is 
affected.  Note, however, that at the level of $\Delta M=0.02$ 
between $z=0$ and 1.7 the cosmology fit is not strongly affected; 
e.g.\ the quadrature sum of dispersion and bias is increased by less 
than 14\% relative to no magnitude evolution. 

If instead we treat possible evolution as an extra fit parameter 
$\alpha$, this removes the bias, replacing it with an increased 
dispersion.  With 
a gaussian prior of 0.04 mag on $\Delta M(z=1.7)$, the parameter 
estimation uncertainty on $w_0$ increases by 4\%, but on $w_p$ 
increases by 41\%, relative to the no evolution case (the dispersion 
in $w_a$ increases by 1\% and in $\om$ by 49\%).  Of course, exchanging 
a systematic bias for a new fit parameter and increased dispersion only 
works if one knows the functional form of the systematic -- that is the 
whole basis of ``self-calibration''.

\subsection{Modified gravity} \label{sec:gamma}

The expansion history is specified by the equation of state ratio 
$w(a)$, but the history of growth of structure in the universe depends 
on both the expansion history and the theory of gravity.  If gravity 
is modified from general relativity, then the observables involving 
growth will be offset from the general relativity predictions.  This 
in turn will cause a bias in cosmological parameters interpreted within 
the framework of Einstein gravity. 

Effects from modification of gravity can be quite complex, and there 
is no general treatment even in the linear regime of structure formation. 
One approach is the growth index parametrization of \cite{groexp}; this 
has been shown highly accurate in the linear growth factor for taking 
into account expansion history effects, and modified gravity in the DGP 
\cite{dgp} braneworld model.  Another interesting approach, not yet fully 
developed, is phenomenological modification of the Poisson equation 
giving the source term of the growth equation; this 
will give a scale dependent growth factor \cite{stabenaujain}. 

In the growth index formalism, the linear growth factor of matter density 
perturbations $g(a)=(\delta\rho/\rho)/a$ is given by 
\beq
g(a)=e^{\int_0^a da/a\,[\om(a)^\gamma -1]}, 
\eeq 
where $\gamma$ is the growth index.  For a matter plus cosmological 
constant universe, $\gamma=0.55$, and the formula is accurate to better 
than 0.05\%.  Modification of gravity will shift $\gamma$; for example 
the DGP growth factor is well fit (to 0.2\%) by $\gamma=0.68$. 

\begin{figure}[!hbt]
\begin{center}
\psfig{file=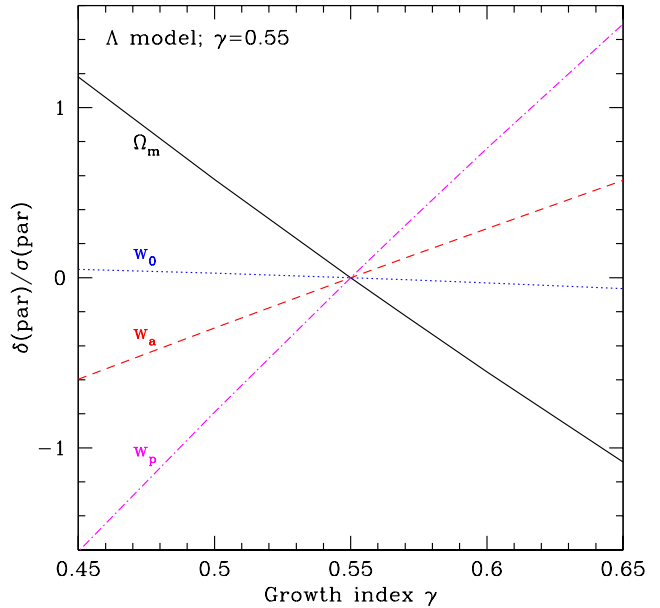,width=3.4in}
\caption{Modification of gravity, if interpreted in an Einstein 
gravity context, will bias the cosmological parameter estimation. 
Here we consider observational constraints from next generation linear 
growth factor measurements to 2\% over $z=0-2.8$, plus supernova and 
CMB measurements; the modified gravity is parametrized through the 
growth index $\gamma$ \cite{groexp}, with $\gamma=0.55$ in general 
relativity.  The $w_0$ and $w_a$ parameters are reasonably robust,
but the sweet spot, or pivot, value $w_p$ (and to a less extent the 
matter density $\om$) can be appreciably biased.
}
\label{fig:biasg}
\end{center}
\end{figure}

To propagate the effects of modified gravity on growth through to 
cosmological parameter bias we offset $\gamma$ from the general relativity 
value.  We then add future measurements giving the linear growth factor 
at $z=0$, 0.4,\dots 2.8 to 2\% precision to the previously considered 
supernova and CMB distance observations.  As shown in Fig.~\ref{fig:biasg}, 
again $w_p$ is strongly biased, while $w_0$ is scarcely affected.  
Deviations in $w_a$ are minor, and in $\om$ are appreciable but not 
severe for $|\Delta\gamma|<0.05$.

\subsection{Density parametrization and crossover bias} \label{sec:dens}

We have emphasized that parametrizations should be robust, i.e.\ unbiased. 
This means that they must be crafted specifically for the questions 
we want answered.  If a parameter $P$ is robust, i.e.\ its fit value 
equals the expectation value, then $P^2$ or more generally some nonlinear 
function $f(P)$ is not unbiased.  Thus if we want to learn about the 
dynamics $w'$--$w$ of dark energy, we should parametrize the equation 
of state directly. 

Using other quantities such as dark energy density $\rho$ or Hubble 
parameter $H$ as a basis for parametrization is known to be improper 
and dangerous, when the aim is the dark energy equation of state. 
For example, \cite{jonsson} clearly demonstrated the instability and 
bias in $w(a)$ from adopting a functional form for $\rho(z)$ or $H(z)$, 
and \cite{recon} showed that binned values $\rho(z_i)$ had similar 
pathology (although \cite{recon} pointed out that binned values in $a$ 
or $\ln(1+z)$ are subject only to the usual, still worrying, 
numerical instabilities).  

Instabilities often lead to crossover behavior, where spurious 
evolution, e.g.\ from $w<-1$ to $w>-1$ occurs.  \cite{jonsson} showed 
instances where this necessarily happened close to the minimum variance 
location -- i.e.\ $w_p\approx-1$ was forced.  Perhaps the earliest 
crossover models were given by \cite{lingrav} (the crossing of $w=-1$ 
is now sometimes called the ``phantom divide'' \cite{hu04}) and have 
some interesting properties relevant to parameter bias.  When 
implemented by two scalar fields, 
\beq 
w(z)=w_1\frac{\delta H_1^2}{\delta H_1^2+\delta H_2^2} + 
w_2\frac{\delta H_2^2}{\delta H_1^2+\delta H_2^2}, 
\eeq 
where $\delta H_i^2$ is the contribution of component $i$ to the 
Hubble expansion equation, the dependence of $w(z)$ on the parameters 
$w_i$ and $w_j$ becomes nonlinear, and errors in their determination 
lead to the error in the reconstructed $w(z)$ running away as $\ln(1+z)$.  
This can even 
lead to false crossover behavior despite both fields having $w>-1$, say. 
Because observations tell us that the averaged value of $w$ is not 
too different from -1, this means that any crossover must have happened 
at reasonably low redshift -- i.e.\ the crossover redshift must be 
fairly close to the pivot redshift, forcing $w_p\approx-1$. 

(Note also that observational systematics as in \S\ref{sec:obsevo} can 
bias $w_0$ one way and $w_a$ the other way, possibly causing a crossover. 
A simple model where the local Hubble flow supernovae, measured 
with different instrumentation than higher redshift supernovae, are 
offset in magnitude from the others can fulfill this condition.  Hence, 
great care must taken with calibration, and a single experiment should 
cover as much of the full redshift range as possible.)

\section{Confidence contours} \label{sec:contour} 

The considerations and calculations in this paper point out that the 
most precisely determined 
value of $w(z)$ -- i.e.\ $w_p$ -- can also be the most sensitive to bias 
and misinterpretation of the nature of the dark energy from 
improper parametrization or observational systematics. 

The presence of bias does not herald failure for understanding dark energy 
but rather advocates for caution in interpretation and for complete 
description of the parameter estimation.  As shown, robust 
parametrization can obviate many of the problems, and, as in 
\S\ref{sec:obsevo}, one can cure bias by judiciously chosen 
additional fit parameters. 

One should also distinguish between the absolute parameter bias $\delta p$ 
and the fractional bias $\delta p/\sigma(p)$; in many cases the fractional 
bias on $w_p$ is larger than on $w_0$ because $w_p$ is more precisely 
determined.  Still, if $w_p$ is used to advertise precision, the fractional 
bias tells to what extent this is an accurate result.  Statisticians 
often combine the precision, or dispersion $\sigma(p)$, with the bias 
$\delta p$ to form the ``risk'' on a parameter, 
\beq 
r(p)=\sqrt{\sigma^2(p)+\delta p^2}=\sigma(p)\,
\left(1+[\delta p/\sigma(p)]^2\right)^{1/2}. \label{eq:risk} 
\eeq 
We see that the fractional bias then gives the increase from the 
statistical dispersion to the parameter risk.  For many of the 
cases discussed in this paper, despite a higher fractional bias 
than $w_0$, $w_p$ still has a lower overall risk.

Risk, however, is a matter of personal comfort not rigorous mathematics, 
in cosmology as in the stock market\footnote{It is true that there are 
rigorous inequalities involving bias.  The Rao-Cram{\'e}r-Frechet 
bound \cite{cramer} states that the variance of a parameter $p$, 
about a possibly biased mean, is 
\beq 
\langle (p-\langle p\rangle)^2\rangle\ge (1+\partial \delta p/\partial 
p)^2/F_{pp}\,. 
\eeq 
This generalizes the usual Fisher result that the variance has a lower 
limit, 
$\sigma_p\ge1/\sqrt{F_{pp}}$.}.  Rather than having statistical 
precision and bias contribute equally, as in Eq.~(\ref{eq:risk}), we 
might be willing to pay a 
premium to have great confidence in our understanding of fundamental 
physics, and weight robustness and low bias more strongly than mere 
statistical precision.  If we want to avoid false conclusions about, say, 
whether something other than the 
cosmological constant pervades our universe, we should seek experiments 
and parametrizations that stringently control bias. 

In any case, quoting a single parameter's 
statistical precision, bias, or risk withholds important information 
due to correlations between parameters. 
The likelihood surface for the set of 
parameters (usually shown as a two parameter confidence contour, 
marginalized or minimized over the other parameters) retains and 
illustrates information on dispersion, bias, and degeneracies. 

\begin{figure}[!hbt]
\begin{center}
\psfig{file=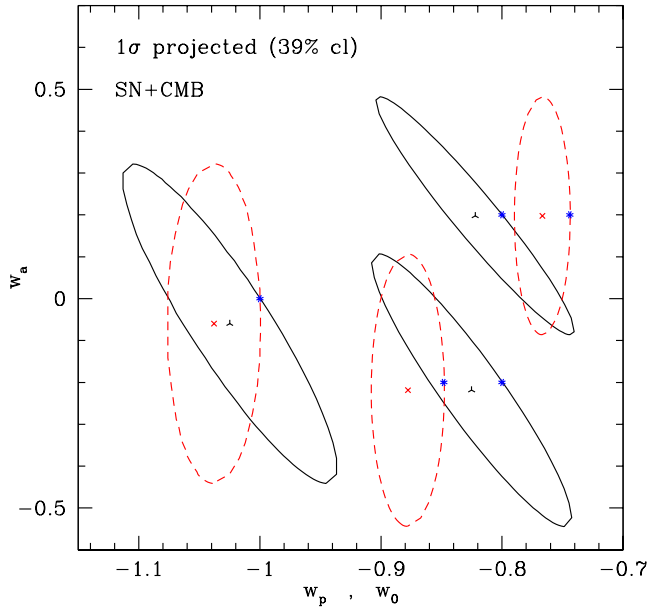,width=3.4in}
\caption{Likelihood contours provide a fairer depiction of bias than 
one dimensional parameter uncertainties.  Various fiducial 
models (with their $w_0$ and $w_p$ values indicated simultaneously 
by blue stars) are biased to new values of $w_0$, $w_p$ by differing 
amounts (respectively shown by black, three legs and red x's), but 
remain at the same confidence level.  Contours in $w_p$--$w_a$ 
space are always vertical, and their degeneracy axis intersects the 
degeneracy axis of the tilted $w_0$--$w_a$ contours at $w_a=0$ (and 
the contour widths for $w_a=0$ are equal in either parametrization). 
}
\label{fig:biasmaw0wp}
\end{center}
\end{figure}

Figure~\ref{fig:biasmaw0wp} illustrates the confidence contours 
for various fiducial models in the case of \S\ref{sec:obsevo} where an 
observational offset in supernova magnitudes as a function of redshift 
biases the cosmological parameters.  To show the bias clearly, we adopt 
$\Delta M(z=1.7)$ such that $w_p$ is biased by $1\sigma$ in each case, 
and plot the $1\sigma$ projected confidence 
contour (39\% confidence level), so the biases can be read off directly 
from projection to the axes.  We simultaneously show the $w_0$--$w_a$ 
and $w_p$--$w_a$ planes: the solid, black contours are for the $w_0$ 
cases, and the three legged symbol shows the best fit, while the dashed, 
red contours are for the $w_p$ cases and the x shows the best fit.  
Each true cosmological model is indicated by the blue stars, one for 
$w_0$, one for $w_p$.  The measurement offset biases the 
parameter estimation such that $w_0$ is much less biased in the 
fractional sense (and in this case in the absolute sense as well).   
However, the true model lies at an equal confidence level in either 
parametrization.

In terms of the individual parameters, then, $w_0$ tends to be more 
robust than $w_p$ for the fractional bias, and one should be wary of 
treating quotes of statistical precision of $w_p$ as robust.  Better 
is to show the joint likelihood surface or confidence contour, which does 
provide an even handed assessment of bias.  We saw in \S\ref{sec:merit} 
that the 
physical incisiveness of the contours depends on both the parametrization 
and location in phase space.  The vertically oriented contours of 
$w_p$--$w_a$ are optimal only in the ``dead zone'' between the 
physically motivated freezing and thawing behaviors (see 
\cite{caldlin,paths}).  Similarly, the area of the contour, 
equal in either plane \cite{hutturaip} and 
conveniently written in the gaussian approximation as 
$\sigma(w_p)\times \sigma(w_a)$, 
is the optimal figure of merit only in the 
Snarkian\footnote{In Lewis Carroll's {\it Hunting of the Snark\/}, in 
searching for the snark, whose nature was completely uncertain, the hunters 
were given a blank map.  
Could Carroll, a mathematician, 
have been commenting on likelihoods and optimal parametrization?} sense.

\section{Conclusion} \label{sec:concl} 

Understanding the new physics lying behind the acceleration of the cosmic 
expansion requires precision in the measurements, accuracy in 
the measurements, and robustness in the interpretation of the data. 
We have tested the concept of the two parameter dark energy equation of 
state phase plane and demonstrated further that the standard $w_0$--$w_a$ 
parametrization 
provides a robust framework for interpreting dark energy, considering 
systematic errors in the observations, more complex dark energy models, 
and gravity beyond general relativity.  The pivot, or decorrelation, 
value of the equation of state, $w_p$, is more subject to bias from 
such effects. 

Seeing the true nature of dark energy, not a subjective, biased 
interpretation, is an exciting challenge.  Robust parametrization, 
yielding a fair, unbiased answer is one step, and points the way to 
another: complementarity of probes plays a crucial role.  We saw at 
the end of \S\ref{sec:biase} that a sufficiently rapid transition 
in the dark energy behavior could lead not only to a bias, but one 
that in a single parameter acts opposite to the direction of the 
transition, spoofing the nature of the dark energy.  Combining more 
than one cosmological probe immunizes against this, by reducing the 
degeneracies that allow such an extreme distortion -- as well as 
guarding against systematic errors (or beyond Einstein gravity) that 
can also bias results. 

We found that one must be wary of uncertainties quoted on individual 
parameters, such as $w_p$ that anyway lacks a standalone physical 
basis, varying based on the probe, survey, priors, and fiducial model. 
The likelihood, or confidence contours in the equation of state plane, 
is a fairer depiction of the consistency of models with data.  A single 
optimal figure of merit cannot be defined, however, for the dark energy 
phase space, but is informed by the physical structure of the dynamics, 
e.g.\ thawing and freezing regions.  The use of the area of contour 
ellipses as proposed by the Dark Energy Task Force is ok, though limited, 
but one should not overinterpret the role of $w_p$ for calculating the 
area as promoting $w_p$ to a favored parameter, due to its flaws. 

In fact, we seem led to either custom figures of merit following the 
physics in the robust $w_0$--$w_a$ parametrization or to taking 
another look at defining a series of benchmark models.  Either way, 
the firm foundation of the dynamics $w'$--$w$ and the standard 
parametrization provides realistic hope for learning the true nature 
of the cosmic acceleration.

\section*{Acknowledgments} 

This work has been supported in part by the Director, Office of Science,
Department of Energy under grant DE-AC02-05CH11231.


\begin{thebibliography}{99}

\bibitem{detf} 
  http://www.nsf.gov/mps/ast/detf.jsp ; \\ 
  https://www.darkenergysurvey.org/the-project/news/ AAAC\_presentation

\bibitem{hutturaip} 
  D.\ Huterer \& M.S.\ Turner, AIP Conf.\ Proc.\ 599, 140 (2001) 
  [astro-ph/0006419]

\bibitem{caldlin}
  R.R.\ Caldwell \& E.V.\ Linder, Phys.\ Rev.\ Lett.\ 95, 141301 (2005) 
  [astro-ph/0505494] 

\bibitem{paths}
  E.V.\ Linder, Phys.\ Rev.\ D 73, 063010 (2006) [astro-ph/0601052]

\bibitem{spergel06} 
  D.N.\ Spergel et al., astro-ph/0603449

\bibitem{coorayhut}
  A.\ Cooray, D.\ Huterer, D.\ Baumann, Phys.\ Rev.\ D 69, 027301 (2004) 
  [astro-ph/0304268] 

\bibitem{linsl} 
  E.V.\ Linder, Phys.\ Rev.\ D 70, 043534 (2004) [astro-ph/0401433]

\bibitem{linprl}
  E.V.\ Linder, Phys.\ Rev.\ Lett.\ 90, 091301 (2003) [astro-ph/0208512] ; 
  E.V.\ Linder, in Identification of Dark Matter (IDM2002), eds.\ 
  N.J.C.\ Spooner \& V.\ Kudryavtsev (World Scientific: 2003) 
  [astro-ph/0210217] 

\bibitem{linhut05}
  E.V.\ Linder \& D.\ Huterer, Phys.\ Rev.\ D 72, 043509 (2005) 
  [astro-ph/0505330]

\bibitem{astier} 
  P.\ Astier, Phys.\ Lett.\ B 500, 8 (2001) [astro-ph/0008306]

\bibitem{huttur} 
  D.\ Huterer \& M.S.\ Turner, Phys.\ Rev.\ D 64, 123527 (2001) 
  [astro-ph/0012510]

\bibitem{welal}
  J.\ Weller \& A.\ Albrecht,  Phys.\ Rev.\ D 65, 103512 (2002) 
  [astro-ph/0106079]

\bibitem{snap}
  http://snap.lbl.gov ; \\ 
  G.\ Aldering et al., astro-ph/0405232

\bibitem{planck}
  http://planck.esa.int 

\bibitem{maor}
  I.\ Maor, R.\ Brustein, J.\ McMahon, P.J.\ Steinhardt, Phys.\ Rev.\ D 65, 
  123003 (2002) [astro-ph/0112526]

\bibitem{simpsonbridle} 
  F.\ Simpson \& S.\ Bridle, Phys.\ Rev.\ D 73, 083001 (2006) 
  [astro-ph/0602213]

\bibitem{klmm}
  A.G.\ Kim, E.V.\ Linder, R.\ Miquel, N.\ Mostek, MNRAS 347, 909 (2004) 
  [astro-ph/0304509]

\bibitem{holzlinder}
  D.E.\ Holz \& E.V.\ Linder, ApJ 631, 678 (2005) [astro-ph/0412173]

\bibitem{linmiq}
  E.V.\ Linder \& R. Miquel, Phys. Rev.\ D 70, 123516 (2004) 
  [astro-ph/0409411]

\bibitem{lintaup}
  E.V.\ Linder, plenary talk at TAUP2005 -- 
  http://supernova.lbl.gov/\~{}evlinder/taup.ppt

\bibitem{huttak}
  D.\ Huterer \& M.\ Takada, Astropart.\ Phys.\ 23, 369 (2005) 
  [astro-ph/0412142]

\bibitem{mohr}
  J.J.\ Mohr, B.\ O'Shea, A.E.\ Evrard, J.\ Bialek, Z.\ Haiman, 
  Nucl.\ Phys.\ B Proc.\ Suppl.\ 124, 63 (2003) [astro-ph/0208102]

\bibitem{groexp} 
  E.V. Linder, Phys. Rev. D 72, 043529 (2005) [astro-ph/0507263]

\bibitem{dgp}
  G. Dvali, G.\ Gabadadze, M.\ Porrati, Phys.\ Lett.\ B 485, 208 (2000) 
  [hep-th/0005016]; 
  C.\ Deffayet, G.\ Dvali, G.\ Gabadadze, Phys.\ Rev.\ D 65, 044023 (2002) 
  [astro-ph/0105068]

\bibitem{stabenaujain}
  H.F. Stabenau \& B. Jain, astro-ph/0604038

\bibitem{jonsson}
  J. Jonsson, A. Goobar, R. Amanullah, L. Bergstrom, JCAP 09, 007 (2004) 
  [astro-ph/0404468]

\bibitem{recon}
  E.V. Linder, Phys. Rev. D 70, 061302 (2004) [astro-ph/0406189]

\bibitem{lingrav}
  E.V. Linder, Phys. Rev. D 70, 023511 (2004) [astro-ph/0402503]

\bibitem{hu04}
  W.\ Hu, Phys.\ Rev.\ D 71, 047301 (2005) [astro-ph/0410680] 

\bibitem{cramer} 
  H.\ Cram{\'e}r, ``Mathematical methods of statistics'', \S32.3 
  (Princeton U.\ Press: 1946); M.\ Kendall \& A.\ Stuart, ``Advanced 
  Theory of Statistics'', 4th ed., \S17 (Oxford U.\ Press: 1979) 

\end{thebibliography}
\end{document}